\documentclass[pra,aps,floats,letterpaper,floatfix,footinbib,preprintnumbers,twocolumn]{revtex4}
\usepackage{amsmath}
\usepackage{hyperref}
\usepackage{color}
\usepackage[mathenv]{}
\usepackage{graphicx}
\usepackage{natbib}

\begin{document}

\title{Performance of a prototype atomic clock based on {\em lin}$||${\em lin} coherent population trapping resonances
in Rb atomic vapor}

\author{
    Eugeniy E. Mikhailov,
    Travis Horrom,
    Nathan Belcher,
    Irina Novikova\footnote{Corresponding author: inovikova@physics.wm.edus}
    }
\affiliation{The College of William $\&$ Mary, Williamsburg, VA, 23187, USA}

\begin{abstract}
We report on the performance of the first table-top prototype atomic clock based  on coherent population
trapping (CPT) resonances with parallel linearly polarized optical fields ({\em lin}$||${\em lin}
configuration). Our apparatus uses a vertical cavity surface emitting laser (VCSEL) tuned to the D${}_1$ line of
${}^{87}$Rb with current modulation at the ${}^{87}$Rb hyperfine frequency. We demonstrate cancellation of the
first-order light shift by proper choice of rf modulation power, and further improve our prototype clock
stability by optimizing the parameters of the microwave lock loop. Operating in these optimal conditions, we
measured a short-term fractional frequency stability (Allan deviation) $2 \times 10^{-11} \tau^{-1/2}$ for
observation times $1~\mathrm{s} \leq \tau \leq 20$~s. This value is limited by large VCSEL phase noise and
environmental temperature fluctuation. Further improvements in frequency stability should be possible with an
apparatus designed as a dedicated {\em lin}$||${\em lin} CPT resonance clock with environmental impacts
minimized.

\end{abstract}



\date{\today}
\maketitle

\begin{figure}[h]
  \includegraphics[angle=0, width=1.00\columnwidth]{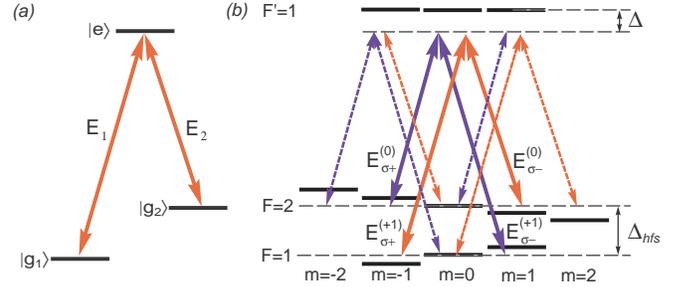}
  \caption{
    \emph{(a)} Idealized three-level $\Lambda$ system that allows for coherent population trapping.
    \emph{(b)} Optical transitions excited during the interaction of two linearly-polarized optical fields at
    the carrier frequency $E^{(0)}$ and the first modulation sideband $E^{(+1)}$ for the
    D${}_1$ line of ${}^{87}$Rb atoms.
    In the presence of a longitudinal magnetic field each light field is decomposed into $\sigma_+$
    and $\sigma_-$ circular components of equal amplitude. Solid arrows indicate the two $\Lambda$ systems responsible for
    magnetic-field insensitive CPT resonances at the hyperfine frequency ({\em lin}$||${\em lin} CPT resonances). One-photon detuning
    $\Delta$ is the frequency difference between the $F=2\rightarrow F'=1$ transition and the unmodulated laser frequency.
    }
  \label{fig:levels}
\end{figure}

In the recent decades impressive progress has been made in development of miniature precision measurement
devices (clocks, magnetometers, gyroscopes, etc.) that use atomic energy levels as a
reference~\cite{vanier05apb,Finland,KnappeAPL04,knappe05oe,litwak04,romalisNature03,romalisPRL05gyro,NISTmagnetometer,BudkerPinesPNAS08,kitchingAPL09}.
A promising scheme for all-optical interrogation of a microwave clock transition in chip-scale atomic devices is
based on the modification of optical properties of an atomic medium under the combined action of multiple resonant
optical fields. For example, under the conditions of coherent population trapping (CPT) simultaneous action of
two optical fields [as shown in Fig.~\ref{fig:levels}(a)] allows ``trapping'' atoms in a non-interacting
coherent superposition of two long-lived hyperfine sublevels of the ground energy state $|g_{1,2}\rangle$ that, under idealized conditions (isolated three-level scheme, no ground-state decoherence), is completely decoupled
from the excited state $|e\rangle$. Such non-interacting state (usually called ``dark state'') exists only when
the differential frequency of two optical fields (two-photon detuning) matches the energy splitting of the
hyperfine states, and leads to a narrow peak in optical transmission - the effect known as electromagnetically
induced transparency, or EIT~\cite{Ref:EITReviewArticle}. The linewidth of a CPT resonance depends on the
intensities of the optical fields, but it is ultimately limited by the finite interaction time of atoms with
light. Since it is possible to obtain CPT resonances as narrow as a few tens to hundreds of
Hz~\cite{erhard01pra,kitchingOL04}, one can lock a microwave oscillator controlling the frequency difference
between two optical fields such that its output frequency is stabilized at the atomic transition frequency
$|g_{1}\rangle-|g_{2}\rangle$~\cite{Cyr}. Frequency stability of such atomic clocks improves for a high-contrast
narrow CPT resonance. Also, minimal sensitivity of the CPT resonance frequency to the experiment environmental
fluctuations (such as temperature, laser frequency and power) is required to ensure long-term stable
operation of the clock.

Alkali metal atoms (Cs, Rb, etc.) are well-suited for practical realization of CPT-based atomic clocks, since
their  ground electronic state consists of two long-lived hyperfine states. Also, their optical transitions are
easily addressable with diode lasers, allowing potential miniaturization of such clock devices. However, the
complex Zeeman structure of these atoms poses several challenges. Only the frequency between magnetic
field-insensitive Zeeman sublevels $m_F=0$ (``clock transition'') of each ground state should be measured to
avoid the detrimental effects of ambient magnetic fields, since good magnetic shielding is not likely possible
in a chip-scale atomic clock. At the same time, for a traditional CPT configuration using two circularly
polarized optical fields, most atoms are concentrated at the magnetic sublevels with the highest angular
momentum $m_F=\pm F$~\cite{jauPRL04} (``pocket'' states). As a result only a small fraction of atoms contribute
to enhancing transmission at the clock transition, leading to very low CPT contrast, and consequently to a
limited clock stability. Various groups proposed a number of techniques to improve the CPT resonance
characteristics~\cite{TaichenachevJETP04,JauPRL04a,ZanonPRL05,shahOL07}, although many of them add to the
complexity of the experimental setup by requiring, for example, two optical fields of different polarizations.

Recently, a promising approach to produce high-contrast CPT resonances with a single phase-modulated laser was
proposed~\cite{TYlinlin05}, taking advantage of a unique level combination of the alkali atoms with nuclear spin
$I$=3/2. In general, two optical fields of the same linear polarization ({\em lin}$||${\em lin} configuration)
do not create a dark state required for a CPT resonance between the $m_F=0$ sublevels  due to destructive
interference of the involved $\Lambda$ systems. However, when two ground states with  total angular momenta
$F$=1,2 are coupled only with an excited state with total angular momentum $F'$=1, a high-contrast
magneto-insensitive CPT resonance can be formed using $m_F=\pm 1$ Zeeman
sublevels~\cite{matisovPRA05,zibrovJETPLett05,matisovPRA09,linlinPRApreprint}. In a vapor cell this situation is
realized only for the D$_1$ line of $^{87}$Rb, when the exited-state hyperfine levels are spectrally resolved.

The relevant $\Lambda$ systems formed by the circularly polarized components of two linearly polarized optical
fields are shown in Fig.~\ref{fig:levels}(b). In the linear Zeeman effect approximation, shifts of $|F=1,
m_F=\pm 1\rangle$ and $|F=2, m_F=\mp1\rangle$ are almost identical. The two-photon CPT resonance  for a
$\Lambda$ system formed by the $\sigma_+$ component of one optical field (e.g., $E^{(0)}_+$) applied to the
$|F=2, m_F=-1\rangle \rightarrow |F'=1, m_{F'}=0\rangle$ transition and the $\sigma_-$ component of the other
optical field ($E^{(+1)}_-$) applied to the $|F=2, m_F=-1\rangle \rightarrow |F'=1, m_{F'}=0\rangle$ occurs at
the unshifted clock frequency (i.e., at the hyperfine splitting $\Delta_{\mathrm{hfs}}$). The same is true for a
symmetric $\Lambda$ system formed by the opposite circular components. The advantage of the
\emph{lin}$||$\emph{lin} scheme over a traditional \emph{circ}$||$\emph{circ} is the absence of any ``pocket''
states, and hence higher contrast of a CPT resonance~\cite{TYlinlin05,linlinPRApreprint}.

In this manuscript we report the first experimental realization of a table-top atomic clock prototype based on a
{\em lin}$||${\em lin} CPT resonance in ${}^{87}$Rb using a VCSEL laser with direct current modulation. Our
system is potentially scalable to miniature applications. We demonstrate the cancellation of the first-order
light shift of a CPT resonance, confirming the previous results, obtained with an externally phase-modulated
narrow-band diode laser~\cite{linlinPRApreprint}. Operating at such light-shift cancellation conditions, we
observed promising short-term frequency stability ($\approx 2 \times 10^{-11}~\tau^{-1/2}$). The long-term
stability is limited by insufficient temperature control of the environment. We expect superior frequency
stability will be possible in a small {\em lin}$||${\em lin} CPT clock designed for good thermal control, low
phase noise, etc.

\begin{figure}[h]
  \includegraphics[angle=0, width=1.00\columnwidth]{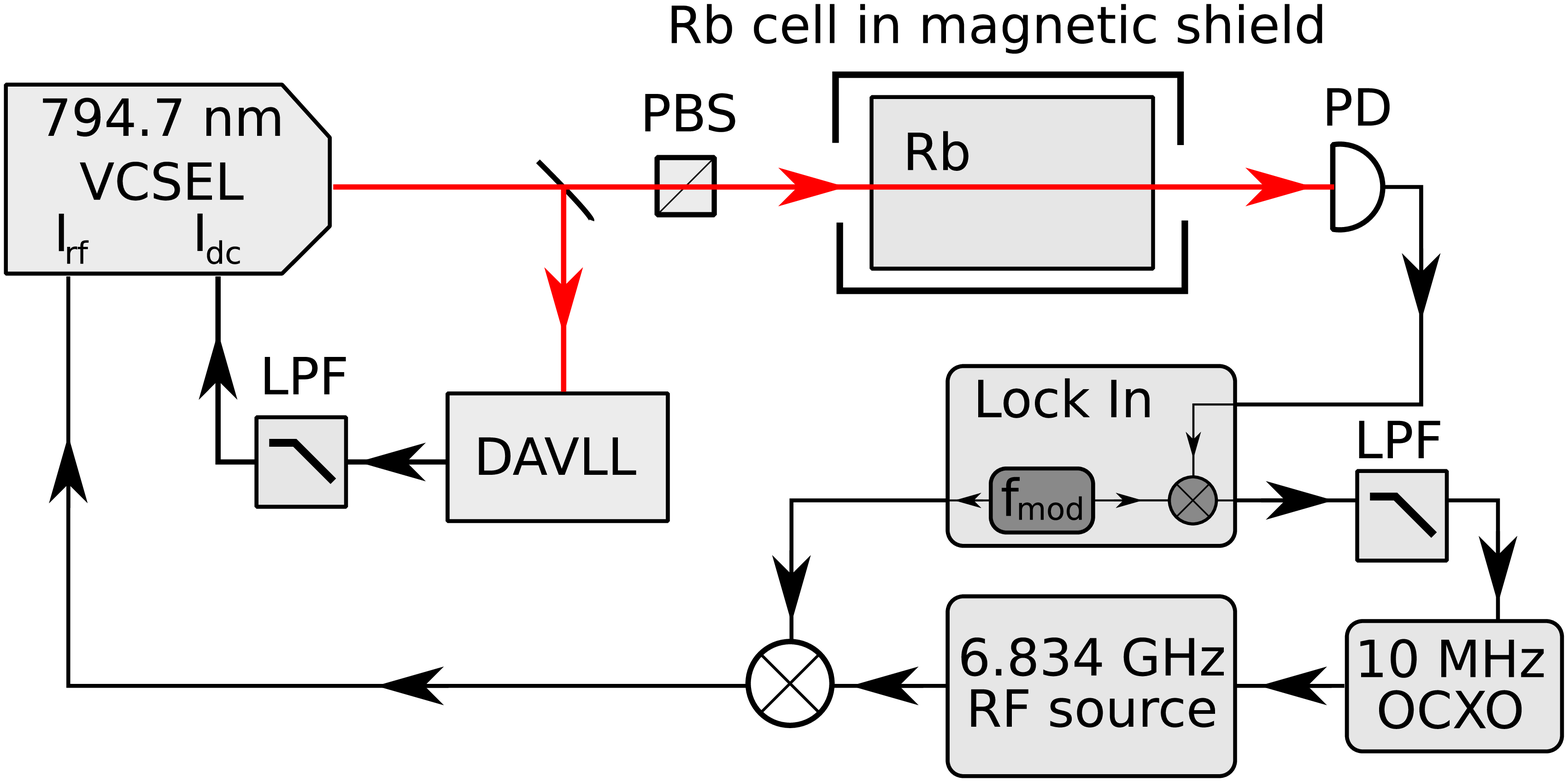}
  \caption{
    Schematic of the experimental setup. Here $PBS$ is a polarizing beam splitter, $LPF$ is a low-pass filter,
    $PD$ is a photodetector, and $VCOCXO$ is a voltage-control oven stabilized oscillator.
    }
  \label{fig:apparatus}
\end{figure}

A schematic of the experimental apparatus is shown in Fig.~\ref{fig:apparatus}. We used a temperature-stabilized
vertical cavity surface-emitting diode laser (VCSEL), operating at $794.7$~nm (Rb D${}_1$ line). The laser
wavelength was locked to the desired atomic transition using a dichroic-atomic-vapor laser lock
(DAVLL)~\cite{yashchukRSI00}. The details of the apparatus design and construction are described in
~\cite{belcherAJP09}. To produce the two optical fields required for CPT, we combined the direct laser current with a
$6.8347$~GHz modulation signal produced by the home-made tunable microwave source described below. For most of
the data described below, the unmodulated laser frequency (carrier) was tuned at or near $5S_{1/2}F=2
\rightarrow 5P_{1/2}F'=1$ transition, while the $+1$ modulation sideband frequency was correspondingly tuned to
$5S_{1/2}F=1 \rightarrow 5P_{1/2}F'=1$ transition. We monitored the intensity ratio between the sideband and the
carrier using a high-finesse Fabry-Perot cavity with free spectral range of approximately 40~GHz (not shown in
the diagram), and we were able to adjust it in a wide range (from zero to more than $100~\%$) by changing the rf
power sent to the VCSEL.

The laser beam with maximum total power 120~$\mu$W and a slightly elliptical Gaussian profile [1.8~mm and 1.4~mm
full width half maximum (FWHM)], was linearly polarized by a polarizing beam splitter (PBS) and then directed
into the cylindrical Pyrex cell (length 75~mm; diameter 22~mm) containing isotopically enriched $^{87}$Rb vapor
and 15~Torr of Ne buffer gas. The cell was mounted inside a three-layer magnetic shielding to reduce stray
magnetic fields, and its temperature was actively stabilized at $47.5^\circ$C. To lift the degeneracy of the
Zeeman sublevels we applied a weak homogeneous longitudinal magnetic field $B\approx 12$~mG produced by a
solenoid mounted inside the innermost magnetic shield. A photodiode (PD) placed after the cell detected the
total transmitted intensity.

\begin{figure}[h]
  \includegraphics[angle=0, width=1.00\columnwidth]{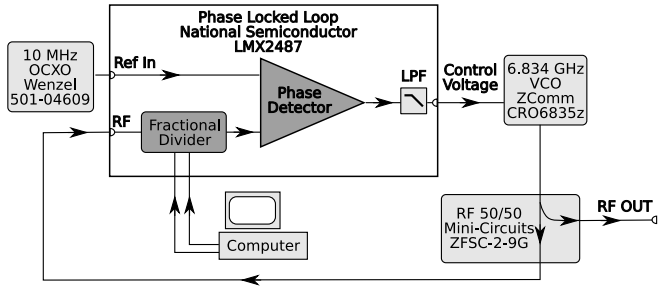}
  \caption{
    Schematic of the 6.835~GHz microwave source.
    }
  \label{fig:rf_source}
\end{figure}

The detailed schematic of the home-made microwave source operating at $6.835$~GHz is shown in
Fig.~\ref{fig:rf_source}. It consists of a Zcomm CRO6835z voltage control oscillator (VCO), for generation of a microwave
field, which is in a phase locked loop (PLL) with a Wenzel 501-04609 voltage controlled oven stabilized $10$~MHz
crystal  oscillator (VCOCXO). A National Semiconductor PLL chip (LMX2487) with a computer controlled fractional
divider allows rough tuning of the microwave frequency with sub Hertz steps in several hundreds mega-Hertz
range, while fine tuning is done via variable voltage of the VCOCXO.

To lock the frequency of the microwave source (and hence the two-photon detuning of the two laser fields) to the
maximum transmission, a slow frequency modulation at $f_m=3$~kHz was superimposed on the $6.835$~GHz microwave
modulation signal. Then the photodetector signal was demodulated at $f_m$ using a lock-in amplifier. The
resulting error signal was fed back to lock the frequency of the $10$~MHz VCOCXO and consequently, the
frequency of the $6.835$~GHz signal was phase-locked to the VCOCXO. The frequency of the locked VCOCXO was measured
by beating it with a reference $10$~MHz signal derived from a commercial atomic frequency standard (SRS FS725).

\begin{figure}[h]
  \includegraphics[angle=-90, width=1.00\columnwidth]{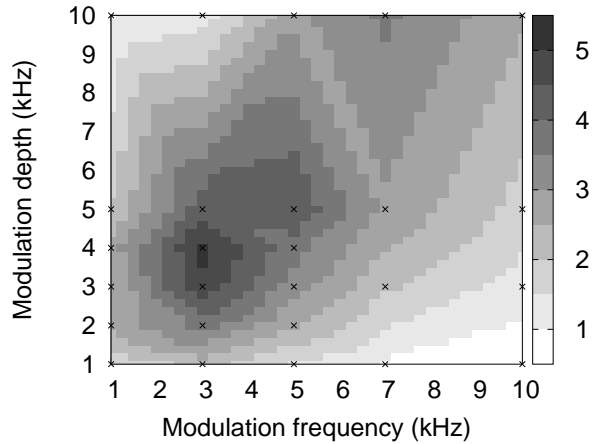}
  \caption{
    Dependence of the clock sensitivity in arbitrary units on the lock-in modulation frequency
    and modulation depth.
    Crosses
    mark  measured data location with the rest of the map recreated via interpolation
    routine.
    }
  \label{fig:lockin_opt}
\end{figure}

To determine the optimal parameters for the microwave lock operation, we measured the error signal as a function
of lock-in frequency and amplitude. The resulting dependence is shown in Fig.~\ref{fig:lockin_opt}. Similar to
previous studies~\cite{BenAroyaOE07} we found that there is a particular combination of the lock-in parameters
that result in the highest slope of the error signal as function of the two-photon detuning: the lock-in
frequency $f_m=3$~kHz and the modulation depth is $4$~kHz. We experimentally confirmed that under these
conditions the microwave lock loop is the most sensitive, and results in the best frequency stability
measurements.


To ensure stable operation of a CPT-based atomic clock, the frequency of a CPT resonance must be maximally
decoupled from any fluctuations of the experimental parameters. For example, any variations in the light
intensity change the measured clock frequency because of the resonance light shift due to interaction of various
VCSEL modulation components with optical transitions. Since the overall CPT resonance shift combines the
contributions of all optical fields on each ground state, it has been previously shown~\cite{linlinPRApreprint}
that careful adjustment of the intensity ratio of two CPT optical fields allows cancellation of the first-order
light shift. Our current measurements confirm that the same cancellation happens for a current-modulated VCSEL
output by adjusting the microwave modulation strength (and hence the sideband/carrier intensity ratio), even
though the VCSEL current modulation is not a pure phase modulation as in the case of electro-optical modulator used in
previous studies.

\begin{figure}[h]
  \includegraphics[angle=0, width=1.00\columnwidth]{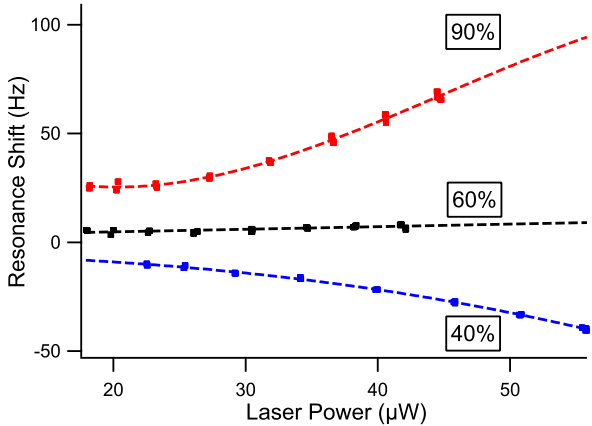}
  \caption{
    Dependence of the measured clock frequency shift on laser power for three different sideband-carrier ratios (40\%,
    60\% and 90\%). For these measurements the carrier laser frequency was tuned by approximately $-200$~MHz from
    $5S_{1/2} F_g=2 \rightarrow 5P_{1/2} F_e=1$ transition.
    }
  \label{fig:shift_vs_power}
\end{figure}

To find the optimal microwave power for light shift cancellation we locked our microwave source on a CPT
resonance. We then monitored changes in the oscillator frequency in response to changing the laser power using an acousical-optical modulator (AOM) before the
Rb cell. Fig.~\ref{fig:shift_vs_power} shows measured CPT light shifts for three different strengths of
microwave modulation, resulting in $40~\%$, $60~\%$ and $90~\%$ sideband/carrier intensity ratios. It is clear
that the resonance light shift reversed its direction when the modulation power was adjusted from highest to lowest, and for a $60\%$ intensity ratio the CPT resonance frequency  was practically independent of the
total laser power. We chose this point to operate our clock to minimize the effect of laser intensity
fluctuations on clock stability.

\begin{figure}[h]
  \includegraphics[angle=0, width=1.00\columnwidth]{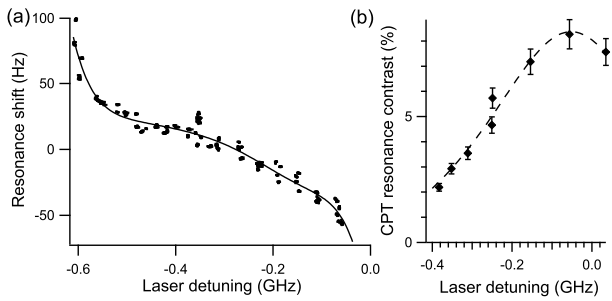}
  \caption{
     Measured clock frequency shift \emph{(a)} and resonance contrast \emph{(b)} as functions of laser detuning
     from the $5S_{1/2} F_g=2 \rightarrow 5P_{1/2} F_e=1$ transition. Total laser power is $60~\mu$W, and the sideband-carrier
    ratio is approximately $50~\%$. Lines are to guide the eye.
        }
  \label{fig:shift_vs_detuning}
\end{figure}

Long-term stability of our clock measurements was limited by dependence of the CPT resonance frequency on a
slowly drifting laser frequency caused by the DAVLL temperature dependence. To minimize this effect, we studied
the CPT resonance shift as a function of laser one-photon detuning $\Delta$. Fig.~\ref{fig:shift_vs_detuning}(a)
shows the change in CPT resonance frequency as the carrier laser frequency was detuned to the red of
$|F=2\rangle \rightarrow |F'=1\rangle$. From this graph it is easy to see that there is no laser detuning with
zero first-order dependence of CPT resonance frequency on $\Delta$. However, a moderate red detuning leads to
weaker dependence of the resonance shift without significantly degrading a CPT resonance contrast [see
Fig.~\ref{fig:shift_vs_detuning}(b)].

\begin{figure}[h]
  \includegraphics[angle=0, width=1.00\columnwidth]{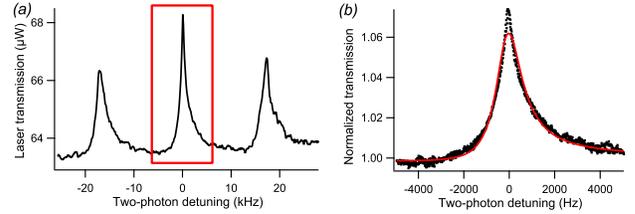}
  \caption{
    \emph{(a)} Measured laser power after the Rb cell as a function of the two-photon
    detuning $\delta-6.834686890$~GHz in the presence of a constant longitudinal
    magnetic field $B=12.3$~mG. The central CPT resonance corresponds to the magnetic field-insensitive
    $\Lambda$ configurations (``clock'' transition), shown with bold solid
    arrows in Fig.~\ref{fig:levels}(b), and two side peaks are
    magnetic field-sensitive CPT resonances. Normalized transmission around the central clock resonance
    is shown (dots) in \emph{(b)} together with a generalized Lorentzian fit
    (solid line).
    Total input laser power is $120~\mu$W, and all the resonances were recorded
    under the optimized light shift cancellation conditions: the sideband-carrier ratio 60\% ,
    the carrier laser frequency $\Delta = -200$~MHz.
    }
  \label{fig:sampleCPT}
\end{figure}

Fig.~\ref{fig:sampleCPT}(a) shows total transmitted power through the Rb cell as the microwave frequency (i.e.,
the two-photon detuning) was scanned around clock transition at the optimal conditions for the first-order light
shift cancellation (sideband/carrier ration $60~\%$, laser detuning $\Delta=-200$~MHz) in the presence of
$\approx 12$~mG longitudinal magnetic field. Three distinct resonances correspond to a magneto-insensitive {\em
lin}$||${\em lin} CPT resonance [central peak, created by the fields depicted by solid arrows in
Fig.~\ref{fig:levels}(b)], and two additional Zeeman-shifted CPT resonance shifts [two side peaks, caused by the
$\Lambda$ configurations shown in dashed arrows in Fig.~\ref{fig:levels}(b)]. As expected from the theory, the
central peak has the highest contrast and the narrowest linewidth: two conditions required for optimal microwave
oscillator locking.

Fig.~\ref{fig:sampleCPT}(b) zooms in on the central {\em lin}$||${\em lin} CPT resonance to analyze its
lineshape.  The resonance was slightly asymmetric due to non-zero one-photon laser
detuning~\cite{taichenachev03pra,mikhailovPRA04}. In this case CPT resonance lineshape should be described by a
generalized Lorentzian function~\cite{knappe02apb}:
\begin{equation}\label{genLor}
T(\delta) =1 + \gamma\,\frac{A\gamma+B(\delta-\delta_0)} {(\gamma)^2+(\delta-\delta_0)^2} \;,
\end{equation}
where $T(\delta)$ is the total laser transmission through the cell normalized to the background $I_{bg}$ (i.e.,
the transmitted power at large two-photon detuning away from CPT resonance), $\delta_0$ is a resonance shift,
$\gamma$ is a CPT resonance linewidth measured at half maximum, and $A$ and $B$ are amplitudes of the symmetric
and anti-symmetric Lorentzian components respectively. All of the above parameters are weakly dependant on
one-photon detunig $\Delta$. Under the conditions of the first-order light shift cancellation that we used in
our atomic clock, the CPT resonance had the following parameters: resonance width $\gamma = 700$~Hz, resonance
contrast $C=6.1$~\% (where the contrast is defined as a ratio between resonance amplitude and background), and
the resonance asymmetry $B/A=0.29$. Fig.~\ref{fig:sampleCPT}(b) shows that Eq.(\ref{genLor}) provides excellent
fit to the experimental lineshape everywhere except for the peak of the resonance, where we observed higher and
sharper transmission than predicted by the fit. This occurs as a result of diffusion of atoms and their repeated
interaction with the laser field~\cite{xiaoPRL06,xiaoOE08}, and it can improve atomic clock frequency stability
by further increasing the overall resonance contrast~\cite{novikovaJMO05}. For example, the measured CPT
contrast exceeded $7$~\% compare to the $6$~\% given by the Lorentzian fit. Also, it is important to note that
the resonant asymmetry is quite small, and may lead to only  a very weak effect of the resonance position on the
lock-in slow modulation parameters~\cite{phillipsJOSAB05}.


The estimated fractional stability of the microwave oscillator locked to the CPT resonance is proportional to
the quality figure $q=C/2{\gamma}$ -- the ratio between the resonance contrast and its full width at half
maximum. The measured resonance parameters ($C=0.07$ and $\gamma=700$~Hz) provide the quality factor $q\approx 5
\cdot 10^{-5}$/Hz. This value implies the fractional frequency stability (Allan variance) at the level of
$\sigma(\tau)\sim 2\cdot 10^{-14}~{\tau}^{1/2}$ if limited only by the photon shot noise~\cite{Vanier03a}:
\begin{equation}
  \sigma(\tau)=\frac{1}{4}\sqrt{\frac{\eta
  e}{I_{bg}}}\frac{1}{q\nu_0} \tau^{-1/2}.
\label{e.stability}
\end{equation}
Here $\nu_0=\Delta_{hfs}=6.834~\mathrm{GHz}$ is the clock reference frequency, $e$ is the electron charge,
$\eta=1.8$~W/A is the photodetector sensitivity (measured optical energy per photoelectron), $I_{bg}$ is the
background intensity, and $\tau$ is the integration time. However, a broad spectral width of VCSEL results in
large residual intensity noise at the output of the cell and therefore significantly degrades realistically
achievable frequency stability.

\begin{figure}[h]
  \includegraphics[angle=0, width=1.00\columnwidth]{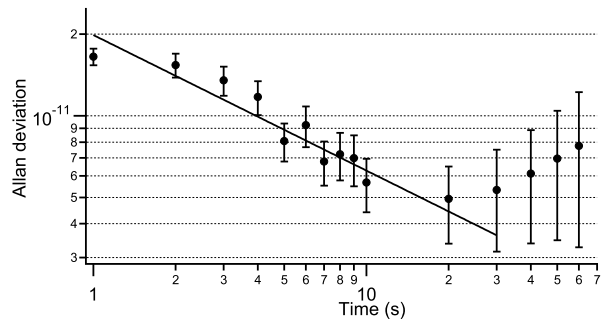}
  \caption{Measured fractional clock stability of the microwave oscillator locked to the CPT resonance.
        }
  \label{fig:stability}
\end{figure}

Fig.~\ref{fig:stability} shows the measured fractional Allan deviation of the clock frequency when our prototype
CPT clock setup operated at optimal light shift cancellation conditions: we detune our carrier by 200~MHz to
decrease sensitivity of the CPT position on the laser detunig (see Fig.~\ref{fig:shift_vs_detuning}), and
maintain 60\% laser field ratio to eliminate light shift dependence (see Fig.~\ref{fig:shift_vs_power}).
The short-term stability was $\simeq 2 \times 10^{-11}~\tau^{-1/2}$ for observation times  $1~\mathrm{s} \leq
\tau \leq 20$~s. This value was most likely limited by the large VCSEL phase noise ($\approx 100$~MHz) as well
as by the stability of our commercial reference clock SRS FS725 with manufacturer fractional stability $<2
\times 10^{-11}$ at 1 second. At longer integration times the stability degraded due to uncontrolled temperature
variations in our table-top apparatus and their effect on the laser wavelength drift that caused the CPT
resonance shift (see Fig.~\ref{fig:shift_vs_detuning}). Despite this non-optimal clock apparatus, the measured
short-term frequency stability is already comparable or better than the values reported for many
recently-demonstrated atomic clocks~\cite{vanier05apb,Finland,knappe05oe,kerncoCPT,novikovaOLcomp}. Our
experimental results also match the theoretically predicted stability limited by broad spectral width of a
VCSEL~\cite{matisovPRA09}.  We expect that both the short- and long-term frequency stability can be further
improved with better temperature stabilization of the experimental apparatus, better laser control, and possibly
using a VCSEL diode with reduced linewidth. 

In summary, we systematically studied a magneto-insensitive CPT resonance in the {\em lin}$||${\em lin}
configuration using a current-modulated VCSEL on the $D_1$ line of ${}^{87}$Rb, and identified the optimal
parameters for atomic clock operation that cancel the effect of the first-order light shift. Employing this
light-shift cancellation in a table-top apparatus (not engineered for stable clock performance), we nonetheless
observed short-term frequency stability of $\simeq 2 \times 10^{-11}~\tau^{-1/2}$ that is comparable to or
better than existing CPT clocks. Significant improvements in such clock frequency stability should be possible
in a small scale device with standard techniques minimizing the impact of the environment.

The authors would like to thank Sergey Zibrov for advice on the experimental apparatus design, and Chris Carlin for helping with the laser lock construction. This research was supported by Jeffress Research grant J-847 and the
National Science Foundation grant PHY-0758010.

\end{document}